\documentclass[sigconf]{acmart}

\AtBeginDocument{%
  \providecommand\BibTeX{{%
    \normalfont B\kern-0.5em{\scshape i\kern-0.25em b}\kern-0.8em\TeX}}}

\copyrightyear{2021}
\acmYear{2021}
\setcopyright{acmlicensed}
\acmConference[CIKM '21] {Proceedings of the 30th ACM International Conference on Information and Knowledge Management}{November 1--5, 2021}{Virtual Event, Australia.}
\acmBooktitle{Proceedings of the 30th ACM Int'l Conf. on Information and Knowledge Management (CIKM '21), November 1--5, 2021, Virtual Event, Australia}
\acmPrice{15.00}
\acmDOI{10.1145/3459637.3482134}
\acmISBN{978-1-4503-8446-9/21/11}

\usepackage{bm}
\usepackage{balance}
\begin{document}
\fancyhead{}
\title{Locate Who You Are: Matching Geo-location to Text for User Identity Linkage}


\author{Jiangli Shao\textsuperscript{1,2}, Yongqing Wang\textsuperscript{1}, Hao Gao\textsuperscript{1,2}, Huawei Shen\textsuperscript{1}, Yangyang Li\textsuperscript{3}, Xueqi Cheng\textsuperscript{1}}

\affiliation{%
	\institution{\textsuperscript{1}Data Intelligence System Research Center, Institute of Computing Technology, Chinese Academy of Sciences\\ \textsuperscript{2}University of Chinese Academy of Sciences\\ \textsuperscript{3}National Engineering Lab for Risk Perception and Prevention, China Academic Electronic and Information Technology}
	\city{ }
	\country{}
}

\email{{shaojiangli19z, wangyongqing, gaohao, shenhuawei, cxq}@ict.ac.cn, liyangyang@cetc.com.cn}

\renewcommand{\shortauthors}{shao et al.}
\def\authors{Jiangli Shao, Yongqing Wang, Hao Gao, Huawei Shen, Yangyang Li, Xueqi Cheng}
\begin{abstract}
Nowadays, users are encouraged to activate across multiple online social networks simultaneously. User identity linkage, which aims to reveal the correspondence among different accounts across networks, has been regarded as a fundamental problem for user profiling, marketing, cybersecurity, and recommendation. Existing methods mainly address the prediction problem by utilizing profile, content, or structural features of users in symmetric ways. However, encouraged by online services, information from different social platforms may also be asymmetric, such as geo-locations and texts. It leads to an emerged challenge in aligning users with asymmetric information across networks. Instead of similarity evaluation applied in previous works, we formalize correlation between geo-locations and texts and propose a novel user identity linkage framework for matching users across networks. Moreover, our model can alleviate the label scarcity problem by introducing external text-location pairs. Experimental results on real-world datasets show that our approach outperforms existing methods and achieves state-of-the-art results.
\end{abstract}



\iffalse
<ccs2012>
<concept>
<concept_id>10002951.10003227.10003236.10003101</concept_id>
<concept_desc>Information systems~Location based services</concept_desc>
<concept_significance>300</concept_significance>
</concept>
<concept>
<concept_id>10002951.10003227.10003351</concept_id>
<concept_desc>Information systems~Data mining</concept_desc>
<concept_significance>300</concept_significance>
</concept>
</ccs2012>
\fi


\maketitle

\section{Introduction}  

Nowadays, the highly developed online techniques make progress in social media applications, encouraging users to share diverse information on different social platforms. User identity linkage (UIL), which aims to reveal the correspondence among different accounts across networks, has been in the spotlight for a long time due to the important role in several downstream tasks, such as user profiling\cite{zhan2017community}, marketing\cite{WangSGC19}, cybersecurity\cite{shu2017user} and recommendation \cite{fan2019metapath}. 

\begin{figure}[!h]
	\centering
	\includegraphics[width=7cm]{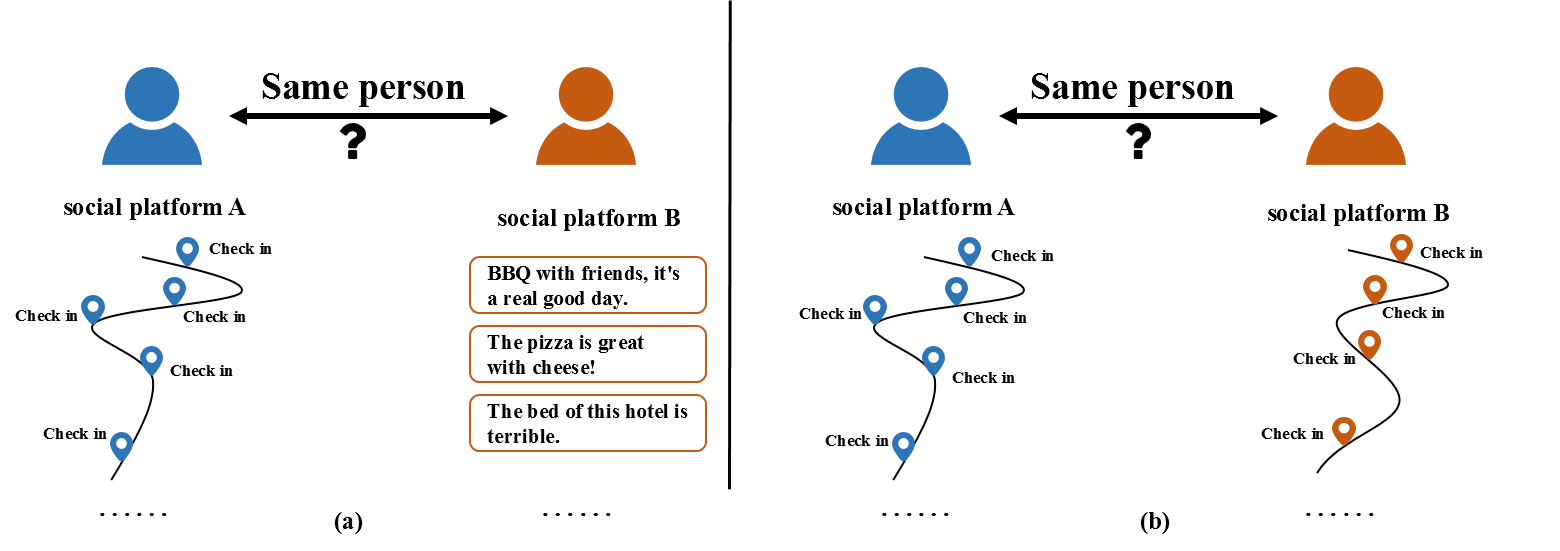}
	\caption{A sketch to show the differences between UIL tasks with asymmetric information(a) and symmetric information(b).}
	\label{asy-anc}
\end{figure} 

Traditionally, most existing approaches in the literature attempt to address the linkage problem with kinds of information, such as profile, content, and structural data\cite{shu2017user,GaoWLSC20}. For instance, researchers extract delicate features from user name, age, gender, registration time, location, and email, and then construct a well-defined prediction model across networks\cite{goga2013exploiting}. Moreover, Nie et al.\cite{nie2016identifying} identified correspondence by learning latent interests from user-generated content. Recently, advances in network embedding techniques drive the progress of UIL modeling by structural features. Some more powerful models have emerged such as PALE\cite{man2016predict}, DeepLink\cite{zhou2018deeplink}, and PAAE\cite{shang2019paae}. Furthermore, by making use of multi-information, LHNE\cite{wang2019user} and TALP\cite{li2020type} attempt to model UIL on heterogeneous networks. Although kinds of information are applied in different modeling, the input data across networks must be symmetric, that is, only those users who have similar types of information in profile, content, and structures can be aligned across networks.  

However, users are also encouraged to post different types of information on different social platforms. For example, a user can travel a scenic spot, post a tweet on Twitter\footnote{https://twitter.com/} and check in on Foursquare\footnote{https://foursquare.com/}. The scenario makes a new challenge for UIL in practice, that is, how to align user identification with so-called "asymmetric information" across networks. Due to the discrepancy of feature spaces, traditional methods may perform worse or even become invalid in this asymmetric situation.

In this paper, we are interested in a common UIL problem based on asymmetric information, i.e., \textit{can we align user identification with texts and geo-locations across network?} To solve the problem, two key issues should be approached ahead: 1) how to evaluate the correlation between texts and geo-locations? 2) how to match pair of users according to sequences generated by texts and geo-locations? Initially, the text-location correlation matrix can be calculated by the co-occurrence between geo-location and corresponding text expression across networks. However, such co-occurrence is hard to be detected especially in the situation with scarce labeled linkage between users. Therefore, external data referring to text-location relationship is also applied in modeling text-location correlation matrix. Based on text-location correlation matrix, during matching stage, a user-user interactive tensor is constructed with user pairs from text sequence to geo-location sequence. On the basis of interactive tensor, a 3D convolution neural network is used to determine whether the pair of two accounts is linked or not. In two real applications, the prediction accuracy resulted by the proposed model is over 89\%, outperforming the state-of-the-art methods modeled by symmetric information. With supplementary of additional text-location relationship from external data, interestingly, the experimental results show a significant improvement, especially when training on datasets with scarce linkage labels. Thus, it shows great potentials in real applications. In a nutshell, the main contributions of our work can be listed as follows:
\begin{itemize}
	\item  A new challenge on utilizing asymmetric information for UIL is defined. In the scenario of aligning users with texts and geo-locations across networks, a novel UIL model is proposed by associating text-location information. To our knowledge, this is the first work to utilize asymmetric information in UIL tasks.	
	\item  To circumvent linkage label sparsity, external data referring to text-location relationship is applied in modeling, leading at most 11\% improvement in datasets with scarce linkage labels.	
	\item  Interestingly, experiments on two real-world dataset show the experimental results from models with asymmetric information outperforms the results from state-of-the-art models with symmetric information, proving great potentials in real applications.
\end{itemize}

\section{Related Work}
Previous works on UIL with heterogeneous information are briefly introduced in this section. With heterogeneous information, Zhang et al.\cite{zhang2014meta} formalize a heterogeneous network and implement random walk to explore same user identities. Recently, the advances in network embedding promote the analysis of heterogeneous information. LHNE\cite{wang2019user} embeds cross-network structural and content information into a unified space by jointly capturing the friend-based and interest-based user co-occurrence in intra-network and inter-network, respectively. Then the learned embedding vectors can be applied in matching user identities. With trajectory data, DPLink\cite{feng2019dplink} solves UIL problem based on heterogeneous mobility data collected from services. However, heterogeneous information applied in those mentioned models still requires to be used symmetrically across networks. Thus, the matching performance is still worse in asymmetric scenario.  

Meanwhile, the idea of interactive tensor is also applied in natural language processing \cite{pang2016text,wan2016deep}. Nevertheless, the proposed interactive tensor, in this paper, concentrates more on exploring correlation between texts and geo-locations. Moreover, the built tensor is sensitive in time, promising better prediction performance in real applications.
\begin{figure}[h]
	\centering
	\includegraphics[width=6cm]{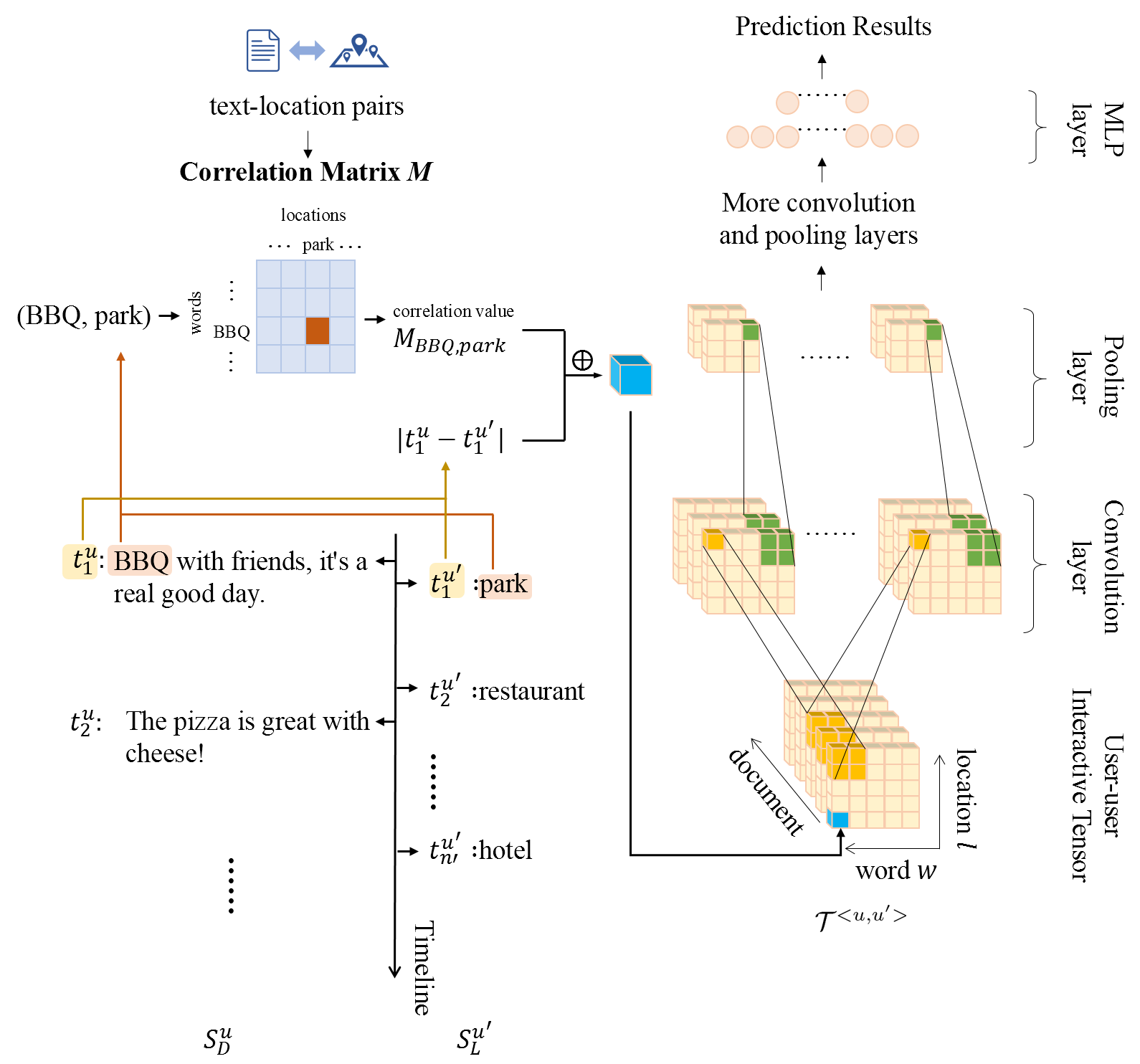}
	\caption{The framework of our model, which takes user record sequences $S_{D}^{u}$ and $S_{L}^{u'}$ as inputs and outputs the probability that user $u$ and $u'$ belong to the same natural person.}
	\label{model}
\end{figure}

\section{Methodology}
In this section, problem statement is introduced at first. Then, the specifies of proposed model are presented. 

\subsection{Problem Formulation}
In the scenario of aligning users with texts and geo-locations across networks, users can check in at spots on one online service $L$ and post corresponding text messages on another online service $D$. Generally, the trajectory of user $u$ can be recorded as chronological sequence $S_L^u=\{(l_k^u, t_k^u)|k=1...N_L\}$, where $l_k^u$ and $t_k^u$ in tuple $(l_k^u, t_k^u)$ refer to the $k$-th location and time generated by user $u$. Meanwhile, posts of user $u$ can be represented as another chronological sequence $S_D^u=\{(d_k^u, t_k^u)|k=1...N_D\}$, where $d_k^u$ and $t_k^u$ refer to the $k$-th document and its publish time generated by user $u$, and document $d_k$ is consisted of word sequence $w_{k1} \ldots w_{km}$. 

For each user identity $u$ in one platform (e.g., serving texting services), the objective of UIL with asymmetric information is finding a mapping function $f$ to identify, if any, its counterpart $u’$ in another platform (e.g., serving location services). The mapping function $f$ can be formalized as, 
\begin{equation*}
	f(\bm{S}_{D}^{u},\bm{S}_{L}^{u'})=\left\{
	\begin{aligned}
		& 1 ,~\text{if $u$ and $u'$ belong to the same person,} \\
		& 0 ,~\text{otherwise.}
	\end{aligned}
	\right.
\end{equation*}

\subsection{Model Framework}
Before diving into the detail of the proposed model, the difference of modeling symmetric and asymmetric information is illustrated firstly. When tackling UIL tasks with symmetric information, similarity can be conveniently evaluated by kinds of metrics defined on distances. However, the pairs of asymmetric data can hardly be evaluated by those distance metrics. Instead of distance metrics, a text-location correlation matrix $\bm{M}$ between words and geo-locations is calculated according to text-location pairs. Given labeled pairs of texts and geo-locations, the correlation value of word $w$ and location $l$ can be defined in a \textit{tf-idf} way, such as:

\begin{equation}
\bm{M}_{w,l}=\log(1+\text{count}(w, l))\cdot \log\left(1+\frac{\sum\limits_{l'\in \bm{L}}^{}\text{count}(w,l')}{\epsilon+\text{count}(w,l)}\right).
\label{eqcorr}
\end{equation}
Here, $\text{count}(w, l)$ denotes the number of occurrences between word $w$ and location $l$ appeared in all documents and their associated location. The $\epsilon$ is a constant.

Based on the calculated correlation matrix $\bm{M}$, a user-user interactive tensor $\mathcal{T}^{<{u,u'}>}$ can be defined as inputs of $S_{D}^{u}$ and $S_{L}^{u'}$. Formally, every element in  $\mathcal{T}^{<{u,u'}>}$ is formulatted as:
\iffalse
\begin{equation}
\begin{aligned}
&\bm{T}_{ijk}^{<{u,u'}>}=(\sum_{c\in d_i}\sum_{l\in L}\sum_{w\in W}\mathbb{I}(w,c)\mathbb{I}(id(c,d_i),k)\mathbb{I}(l,l_{j})\cdot\bm{M}_{w,l})||\\
&\qquad\quad |t_{i}^{u}-t_{j}^{u'}|\\
&1\leqslant i\leqslant |\bm{S_{D}^{u}}|,\ 1\leqslant j\leqslant |S_{L}^{u'}|,\ 1\leqslant k\leqslant K,\ i,j,k\in N^*
\end{aligned}
\end{equation}
\fi
\begin{equation}
\begin{aligned}
&\qquad\quad\mathcal{T}_{ijk}^{<{u,u'}>}=\bm{M}_{w_{ik},l_j} \oplus |t_{i}^{u}-t_{j}^{u'}|.\\
&1\leqslant i\leqslant |S_{D}^{u}|,\ 1\leqslant j\leqslant |S_{L}^{u'}|,\ 1\leqslant k\leqslant K,\ i,j,k\in \mathbb{Z}^{+}.
\end{aligned}
\end{equation}
Here, $\oplus$ represents the concatenation operation and operator $|x|$ means the absolute value of number $x$. Note that the length of document is truncated or padded to $K$ in order to implement the following convolution operation. %The function $R_{\bm{M}}(w,l)$ will retrieve the value related to word $w$ and location $l$ in correlation matrix $\bm{M}$. 

The 3D convolutional operation\cite{7301342} and dynamic pooling\cite{socher2011dynamic} strategy are then used to extract different levels of matching patterns. Let $\bm{z}^{l}$ denote the feature map in $l$-th layer and $\bm{z}^0=\mathcal{T}^{<{u,u'}>}$. Convolution layers and dynamic pooling layers are defined as follows:
\begin{equation}
\bm{C}_{i,j,k}^{(l,n)} = \sigma \left(\sum\limits_{n'=0}^{N_k-1}\sum\limits_{p=0}^{K_p-1}\sum\limits_{q=0}^{K_q-1}\sum\limits_{r=0}^{K_r-1}\bm{w}_{p,q,r}^{(l,n)}\cdot \bm{z}_{i+p,j+q,k+r}^{(l-1,n')}+b^{(l,n)} \right),
\end{equation}
\iffalse
\sum\limits_{n'=0}^{N_k-1}
_{i,j,k}
\fi
\begin{equation}
\bm{z}_{i,j,k}^{(l,n)} =\max\limits_{0\leqslant p< d_l}\max\limits_{0\leqslant q< d_w}\max\limits_{0\leqslant r< d_h}\bm{C}_{i\cdot d_l+p,j\cdot d_w+q,k\cdot d_h+r}^{(l,n)}.
\end{equation}
Here, $\bm{C}^{(l,n)}$ is the convolutional output of $n$-th kernels in $l$-th layers. The $\sigma(\cdot)$ denotes ReLU activation function. The size of the kernel is $K_p*K_q*K_r$ and $N_k$ is the number of channels. Meanwhile, the length, width and height of pooling operator at $l$-th layer is denoted as $d_l$, $d_w$ and $d_h$, determined by the size of $l$-th convolutional feature map $\alpha*\beta*\gamma$ and the size of pooled feature map $\alpha'*\beta'*\gamma'$, i.e., $d_l=\lceil \alpha/\alpha' \rceil,d_w=\lceil \beta/\beta' \rceil,d_h=\lceil \gamma/\gamma' \rceil$

Eventually, We use a MLP (Multi-Layer Perception) to produce the final prediction results:
\begin{equation}
\hat{y} = sigmoid(\bm{W\cdot z}+\bm{b}).
\end{equation}

During training process, the objective function is formulated by cross entropy, i.e., minimizing the loss
\begin{equation}
Loss=-\sum\limits_{i=1}^{N}[y^{(i)}ln(\hat{y}^{(i)})+(1-y^{(i)})ln(1-\hat{y}^{(i)})],
\end{equation}
where $y^{(i)}$ is the label of the $i$-th training instance and $N$ is the number of training samples.

Overall, the framework of the proposed model is briefly depicted in Figure~\ref{model}. 

\section{Experiment}
\textbf{Datasets} The experiments are conducted on two real-world datasets. The first one is the Twitter-Foursquare dataset, which is collected by \cite{zhang2014transferring}. This dataset contains the trajectories data from Foursquare, a popular location-based social network, and Twitter, a micro-blog social network around the world. Null values are left out in this dataset and there remain 2044 linked user accounts in total. The other dataset comes from Dazhong Dianping\footnote{https://www.dianping.com/}, a popular Chinese social application with over ten million registered users. To build this dataset, we collect 2218 users’ review and check-in messages at Dazhong from the year 2012 to 2014 and regard them as two types of input information. Moreover, review and check-in data published by same user at same timestamp are discarded to imitate the behavior of users on two distinct platforms in real world. More details about the two datasets are listed in Table~\ref{tab:data}.
\begin{table}[!h]
	\caption{A brief description of two datasets}
	\label{tab:data}
	\begin{tabular}{ccl}
		\toprule
		&\#users&\#records\\
		\midrule
		Foursquare& 2044& 37061\\
		Twitter& 2044& 4084874\\ 
		\midrule
		Dazhong check-in& 2218&82605\\
		Dazhong review& 2218& 74821\\ \hline
		
	\end{tabular}
\end{table}
\\\textbf{Baselines} We compare our models with the following four baselines. Network structure features are not used in all of the models to ensure the fairness of comparison and hyper-parameters of baselines are the same as those in the original paper.

\textbf{MNA\cite{kong2013inferring}}: This model extracts text features, temporal features, and spatial features from input sequences and sends them to an SVM to finally give the predicted results.

\textbf{STUL\cite{chen2017exploiting}}: STUL captures stay regions and corresponding time distribution features from input sequences and identifies linked user accounts by comparing similarities in these two features.

\textbf{DPLINK and DPLINK*\cite{feng2019dplink}}: DPLINK is an end-to-end deep neural network model that uses LSTM to capture features in the input sequence and use co-attention mechanisms to fusion two trajectories. DPLINK* is a variant of the original DPLINK model in which different types of features are utilized as the input of LSTM.\\
\textbf{Settings}
In the experiments, the text data is cleaned by removing stopwords, punctuation, and URLs. For the Twitter-Foursquare dataset, the length of tweets is limited to 50 and each Foursquare geo-location is converted to a one-hot vector with 476 dimensions according to its category. For the Checkin-Review dataset, the review length is restricted to 100 after word segmentation and each check-in location is converted to a one-hot vector with 45 dimensions according to its category.

The negative sample is constructed by choosing non-linked users from two platforms. The number of positive and negative samples keeps the same. We use 80\% data for training, 10\% for validating, and 10\% for testing. As for the hyperparameter settings, 2 convolutional layers and 2 pooling layers are used in model for both datasets. Kernel size is set to 3*3*3 and the number of channels is 8. Performances of models are evaluated by F1, accuracy(ACC), and AUC metrics.\\
\textbf{Results} Firstly, we make an overall comparison of the prediction performance among all of these models. The results are shown in Table~\ref{res}.
\begin{table}[!th]
	\caption{Performance comparison of different methods. The "Type" column shows the type of information used in UIL tasks in which "sym" denotes symmetric information and "asym" denotes asymmetric information.}
	\label{res}
	\resizebox{\linewidth}{!}{
		\begin{tabular}{l|c|ccc|ccc}
			\hline
			& & \multicolumn{3}{c|}{Twitter-Foursquare} & \multicolumn{3}{c}{Checkin-Review} \\ \hline
			&Type &F1       &ACC       &AUC      & F1      & ACC      &  AUC    \\ \hline
			MNA	&sym& 0.7278      & 0.7415      & 0.7720     & 0.8917      & 0.8919      & 0.9573     \\
			STUL	&sym&0.7563       &0.7683       &0.8573      &0.9099       &0.9099       &0.9605      \\
			DPLINK	&sym&0.8242       &0.8244       &0.8838      &0.9459       &0.9459       &0.9802      \\
			DPLINK*	&asym&0.7826       &0.7854       &0.8788      &0.9030       &0.9032       &0.9716      \\
			OURS	&asym&\textbf{0.8926}       &\textbf{0.8927}       &\textbf{0.9327}      &\textbf{0.9685}       &\textbf{0.9684}       &\textbf{0.9892}\\     
			\hline
	\end{tabular}}
\end{table}
We can notice that our model outperforms the other four baselines on both of the two datasets. The proposed model obtains absolute improvements for about 6.8\% and 2.3\% in terms of F1, 6.8\% and 2.3\% in terms of ACC, and 4.9\% and 0.1\% in terms of AUC when compared with the best baseline on two datasets respectively, demonstrating the effectiveness of the proposed framework in UIL task. According to results from MNA, STUL, DPLINK and our model, traditional models with symmetric information have a poor performance in this asymmetric situation as discrepancies between the text feature space and location feature space. However, our model avoids the similarity comparison between two different feature spaces via computing correlation between words and geo-locations, which might explain the excellent results of our model on this task. Moreover, the proposed model explores more precise matching patterns in word-location level, thus enabling better performance than DPLINK*, which extracts matching patterns in sequence representation level.

\begin{figure}[!htb]
	\centering
	\includegraphics[width=6cm]{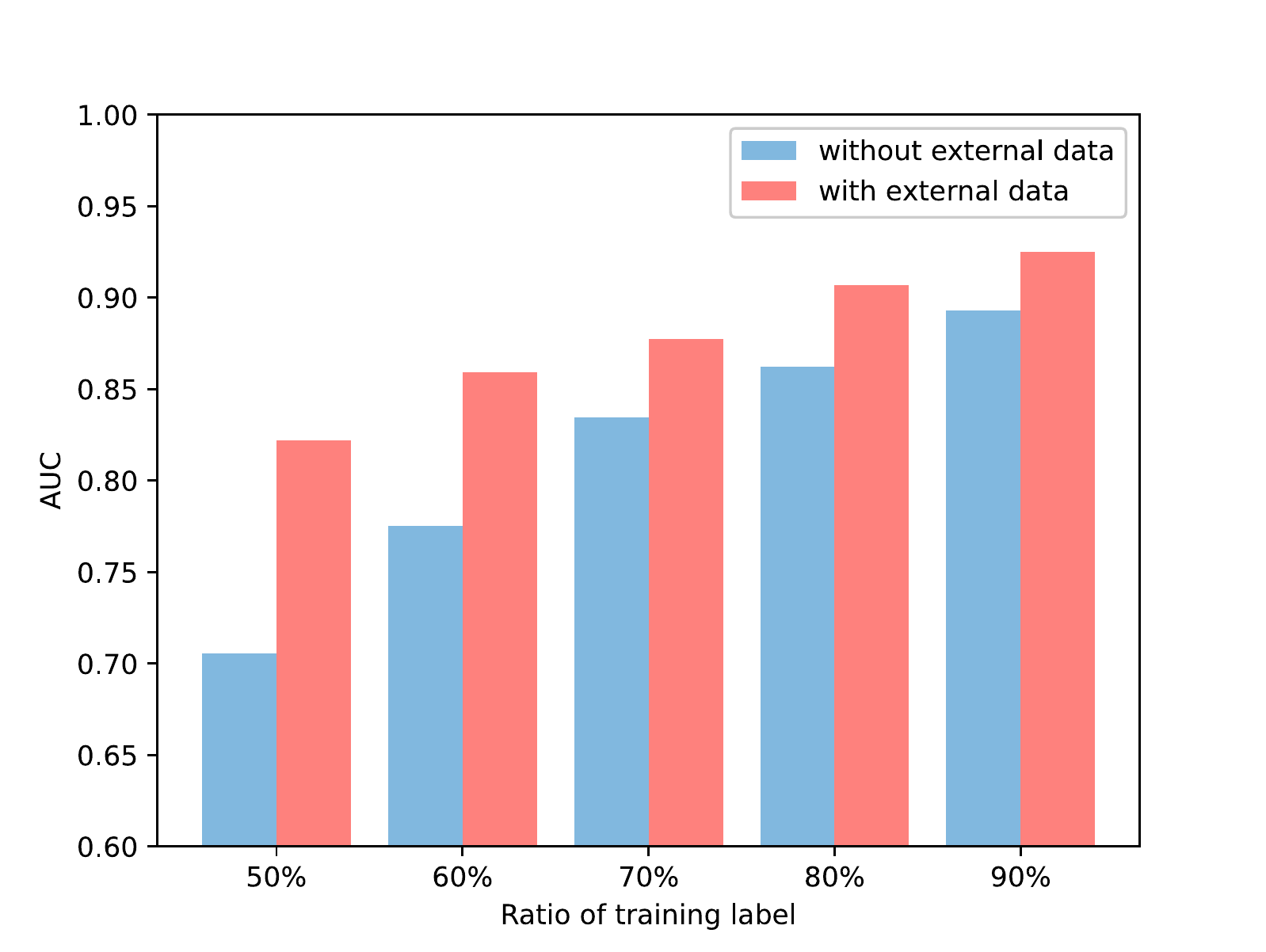}
	\caption{Effects of adding external data to our model over different ratios of training label}
	\Description{AUC curve}
	\label{pic:corr}
\end{figure}

According to the correlation matrix defined in Eq.~\ref{eqcorr}, the proposed correlation matrix can also be calculated by relevant text-location pairs. As a consequence, we introduce the yelp\footnote{A crowd-sourced review site for businesses. Website: https://www.yelp.com} review data and foursquare review data for better estimating the correlation matrix. With the additional text-location pairs, the estimation bias caused by scarce linkage data can be reduced to some extent. As depicted in Figure~\ref{pic:corr}, with the supplementary data, the performance on AUC is increased by 11.7\%, 8.4\%, 4.3\%, 4.5\% and 3.2\% when using 50\%, 60\%, 70\%, 80\% and 90\% training label respectively. Interestingly, according to the improvement on different settings, the introduced data can take more positive effects on scarce linkage labels. It is practical to be applied in real application.

\begin{figure}[!htb]
	\centering
	\includegraphics[height=5cm]{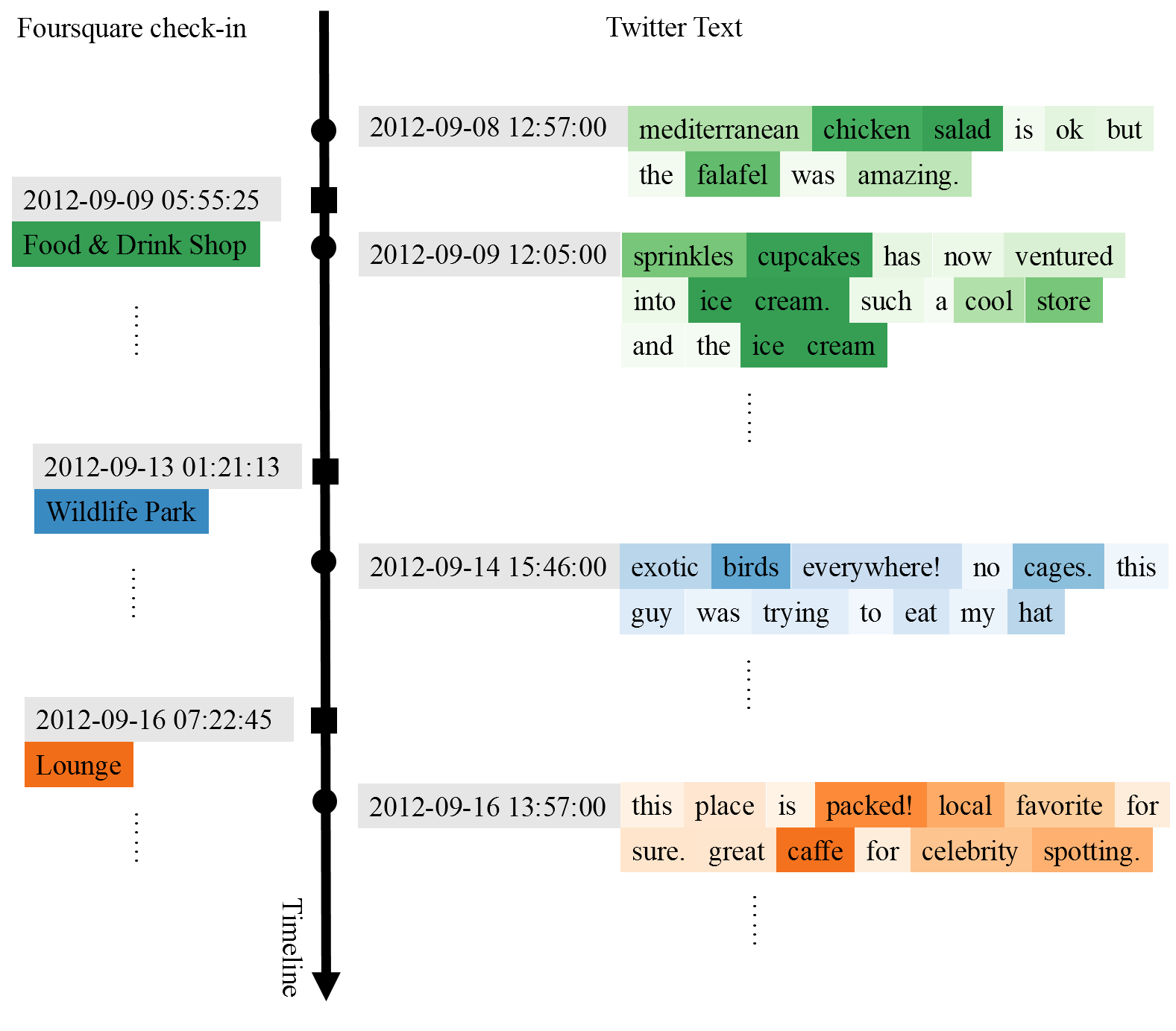}
	\caption{A case to show the results of the proposed model. Squares and circles on the timeline represent the check-in on foursquare and tweets on twitter respectively. Different colors of foursquare check-in mark different locations and the corresponding color of words in tweets indicate the value in correlation matrix. The darker the color, the higher the correlation between words and locations.}
	\Description{case study}
	\label{case}
\end{figure}

Finally, to give a deeper insight into the linkage mechanism of the model, we exhibit a case of explored linked user accounts in test set in figure~\ref{case}. The correlation between words and locations is highlighted in the figure. We can observe that, when visiting some places, users are likely to check in on one platform and post some related tweets on the other at a similar time. In other words, during some periods, records on two platforms show not only temporal but also semantic correlation. The findings are consistent with our intuition and serve as the fundamental matching patterns of the proposed model. By detecting and aggregating such kinds of matching patterns, the framework can correctly discover linked user accounts.

\section{Conclusion}
In this paper, we propose a novel user identity linkage model by matching geo-locations and texts in two social platforms. With inputs of asymmetric information, the proposed model will first construct a user-user interactive tensor via times and the correlation matrix between words and locations. Then, 3D convolutional operations are conducted on the tensor to capture the distinct features. Experimental results show that our model can greatly improve the user identity linkage performance compared with baselines. Meanwhile, we observe that the proposed framework can overcome the training linkage label scarcity problem to some extent by adding additional text-location pairs to calculate the correlation matrix of words and geo-locations, showing great potentials in real applications.

\begin{acks}
This work was funded by the National Natural Science Foundation of China under grant numbers 61802371, 91746301, U1836111, the National Key Research and Development Program of China under grant numbers 2018YFC0825200 and the National Social Science Fund of China under grant number 19ZDA329.
\end{acks}

\bibliographystyle{ACM-Reference-Format}
\balance
\bibliography{ref}


\end{document}